\documentclass[preprint,epsf,aps,floats]{revtex4}
\usepackage{amsmath,amssymb,bm,epsfig,psfrag}
\renewcommand\d{\partial}
\newcommand\p{{\bm{p}}}
\newcommand\q{{\bm{q}}}

\renewcommand\k{{\bm{k}}}
\newcommand\grad{\bm{\nabla}}
\newcommand\+{\dagger}
\newcommand\<{\langle}
\renewcommand\>{\rangle}
\newcommand\epsp{\epsilon_{\p}}

\newcommand{\be}{\begin{equation}}
\newcommand{\ee}{\end{equation}}
\newcommand{\ber}{\begin{eqnarray}}
\newcommand{\eer}{\end{eqnarray}}

\newcommand\beq{\begin{eqnarray}}
\newcommand\eeq{\end{eqnarray}}

\def\Dsl{\,\raise.15ex \hbox{/}\mkern-12.8mu D}

\begin{document}

\title{$\eta/s$ of the Normal Phase of Unitary Fermi Gas from $\varepsilon$ Expansion}
\author{Andrei~Kryjevski}
\affiliation{Department of Physics and Center for Computationally Assisted Science and Technology,~North Dakota State University,~Fargo,~ND~58108}

\begin{abstract}
Using $\varepsilon$-expansion technique we compute $\eta/s$, where $\eta$ is the shear viscosity, $s$ is the entropy density, 
of the normal phase of unitary Fermi gas in $d=4-\varepsilon$ dimensions to LO in $\varepsilon$. We use kinetic theory approach and 
solve transport equations for medium perturbed by a shear hydrodynamic flow. The collision integrals are calculated to $\varepsilon^2$ which is LO. 
The LO result is temperature independent with $\eta/\rm s\simeq (0.11/\varepsilon^2)(\hbar/k_B).$ The $d=3$ prediction for $\eta/\rm s$
exceeds the $\hbar/4 \pi k_B$ bound by a factor of about $1.4.$
\end{abstract}


\maketitle
\section{Introduction}
\label{sec:intro}
Cold dilute gas consisting of two Fermion species (which will be labeled as ${\uparrow},{\downarrow}$) with two body interactions characterized by 
$$
n\,a^3\gg 1,~n\,r_0^3\ll1,
$$
where $n$ is the number density, $a$ is the $s$ wave scattering length and $r_0$ is the effective range of the potential, named ``unitary Fermi gas" is one 
of the simplest examples of a strongly interacting Fermi system. Nevertheless, the unitary Fermi gas is only a slight idealization of dilute neutron matter 
($a \simeq-18 fm,\,r_0 \simeq 2.6 fm$ in the $^1{\rm S}_0$ channel) that may exist in the crust of neutron stars \cite{Bertsch:1999xx,Yakovlev:2004iq}. Also, it has been studied extensively in the experiments on 
trapped cold atomic gases with tunable interactions using the Fesh\-bach resonance technique, {\it e.g.}, 
\cite{PhysRevLett.92.040403,PhysRevLett.92.120401,PhysRevLett.92.120403,PhysRevLett.92.150402,PhysRevLett.93.050401}. In 
these experiments, they have an amazing control over both the strength of the interaction and the composition ({\it i.e.}, imbalance 
of concentrations of pairing Fermi particles, or polarization) of the system (see \cite{bloch-2008-80} for a review of experiments). From the 
theoretical point of view, the unitary Fermi gas lacks any intrinsic scale and is expected to exhibit universal properties. At the same time, a theoretical 
description of unitary Fermi gas is difficult due to the apparent absence of a small dimensionless parameter which could be used in a perturbative expansion, 
thus making the usual perturbative techniques unreliable. 

Properties of these systems have been studied in various Monte Carlo simulations (see \cite{PhysRevA.83.063619} for a compilation of results),
and using various theoretical approaches ranging 
from large-$N$ expansion \cite{veillette:043614} and functional renormalization group \cite{diehl:053627} to phenomenological density functionals \cite{bulgac:040502}. 
See \cite{giorgini:1215} for a review. But an analytical {\it systematically improvable} technique is still needed. 
For example, a Monte Carlo approach is not well-suited to study dynamical quantities (such as dynamical response function) and becomes impractical for the polarized 
case due to the sign problem.
 
Inspired by the ideas of Nussinov and Nussinov \cite{Nussinov:2004nn}, an analytical technique similar to the $\varepsilon$-expansion in the theory of critical 
phenomena \cite{Wilson197475} has been proposed by Nishida and Son \cite{Nishida:2006br}. 
The idea is to describe three dimensional unitary Fermi gas using perturbation theory in the dimensionality of space, $d$, around $d=4$ and $d=2.$ 
The rational for this approach is that unitary Fermi gas simplifies to
\begin{itemize}
\item{non-interacting Fermi gas in $d=2$,} 
\item{the system of non-interacting spin 0 dimers in $d=4$.} 
\end{itemize}
This approach is also applicable to the case of large but finite scattering length (Fermi gas near unitarity) \cite{PhysRevA.75.063617,Chen:2006wx}. 
Predictions for $d=3$ may be obtained by extrapolating the series expansion to $\varepsilon=1,$ possibly employing various convergence improvement techniques, such as 
Pade approximation to the Borel transformation of the series, conformal mapping technique, {\it etc.} \cite{ZinnJ}.

This approach has been used to 
study zero temperature equation of state to the next-to-next-to-leading order (NNLO) in $\varepsilon$ \cite{arnold-2007-75,PhysRevA.79.013627}. For the Bertsch parameter, $\xi,$ 
use of expansions around $d=4$ and $d=2$ with the use of Pad\'e approximants
has given $\xi = 0.36\pm0.02$ \cite{PhysRevA.79.013627}. A recent fixed-node diffusion Monte Carlo simulation has yielded 
$\xi \leq 0.383(1)$ \cite{PhysRevLett.106.235303} and recent lattice calculations extrapolated to zero temperature have produced 
$\xi= 0.37(5)$ \cite{PhysRevA.78.023625} and $\xi=0.3968^{+0.0076}_{-0.0077}$ \cite{Endres:2012cw}.

To NLO in $\varepsilon$ the expansion around $d=4$ has been used to calculate the fermion quasi-particle spectrum \cite{Nishida:2006br,PhysRevA.75.063617} and 
good agreement with Monte Carlo results was found \cite{PhysRevLett.95.060401}.
See \cite{Nishida:2010tm} for a review.
Also, atom-dimer and dimer-dimer scattering in vacuum has been studied \cite{Rupak:2006jj}. The low energy density response function has been calculated \cite{PhysRevA.81.043617}. 
The polarized case has been studied and a polarization induced superfluid-normal phase transition has been found \cite{PhysRevA.75.063617,Rupak:2006et}. Also, effective Lagrangian 
functional (Landau-Ginzburg-like functional) for this system has been derived and used to study the structure of a superfluid vortex, and of the superfluid-normal 
interface in the polarized case, all to NLO \cite{Kryjevski:2007au}. The finite temperature properties of the unitary Fermi gas 
have been studied to NLO. In particular, the critical temperature of the superfluid-normal phase transition has been calculated and good quantitative agreement with numerical results 
has been found \cite{Nishida:2006rp}. 

In all of the above, the NLO corrections have been found to be reasonably small, especially in the non-polarized case. 
All these results suggest that already at the next-to-leading order (NLO) in $\varepsilon$ the expansion around $d=4$ may be a useful tool for the unitary Fermi gas observables,
and that working to a higher order in $\varepsilon$ combined with employing results of expansion around $d=2$ and convergence improvement techniques is expected to produce improvements
in accuracy.

So, as another step in the investigation of the $\varepsilon$-expansion usefulness here we compute shear viscosity of the normal phase of unitary Fermi
gas in $d=4-\varepsilon$ spatial dimensions to LO in $\varepsilon.$ 
Shear viscosity, $\eta$, is a characteristic of the fluid's internal friction.
Determination of this particular transport coefficient has received a lot of attention in recent years, {\it e.g.}, \cite{Cao07012011}. 
See \cite{0034-4885-72-12-126001} for a review. This is, in part, due to the fact that in the strong coupling limit of a large class of 
systems with gauge interactions that can be analyzed from the AdS/CFT correspondence $\eta/s=\hbar/4 \pi k_B,$ where $s$ is the entropy density, 
$k_B$ is the Boltzmann constant. And a lower bound 
\ber
\eta/s \geq \frac{\hbar}{4 \pi k_B}
\label{etasbound}
\eer 
for {\it any substance} has been proposed \cite{Kovtun:2004de}. Both the quark gluon plasma produced 
in the relativistic heavy ion collisions and the unitary Fermi gas have been shown to have \cite{PhysRevC.77.034905,PhysRevLett.99.172301,PhysRevA.76.063618,springerlink:10.1007/s10909-007-9589-1}
\ber
\eta/s \leq 0.5 \frac{\hbar}{k_B}
\label{etasexpt}
\eer
and are considered as the perfect fluid candidates.  
In this work (\ref{etasbound}) will be treated as a benchmark allowing one to determine
usefulness of the $\varepsilon$-expansion technique.

In Section \ref{sec:transporteq} shear viscosity calculation is set up in the $\varepsilon$-expansion framework.
In Section \ref{sec:Cs} calculation of the normal phase collision integrals in $d=4-\varepsilon$ to LO in $\varepsilon$ is presented.
In Section \ref{sec:result} the linearized transport equation solutions for the shear hydrodynamic flow are presented.
Using the previously calculated result for the normal phase pressure LO result for $\eta/s$ is presented which is then compared with the $\eta/s$ predictions
by some other methods. 
\section{Linearized Transport Equations and Expression of the Shear Viscosity}
\label{sec:transporteq}
Due to universality of the unitary Fermi gas, 
we choose to work with local four-Fermi interaction with a coupling constant $c_0$ tuned 
to reproduce infinite scattering length.
The Lagrangian density of
the unitary Fermi gas is 
(here and below $\hbar=k_B=1$)
\ber 
\mathcal{L} = \psi^{\dagger}\left(i\d_t + \frac{\grad^2}{2m} + \mu\right)\psi
 + c_0\,n_\uparrow\,n_\downarrow,
\label{lagrangian}
\eer
where $\psi=(\psi_\uparrow,\psi_\downarrow)^T,$ is spin-1/2 fermion field, and $n_i=\psi^{\+}_i\,\psi_i,$ with
$i={\{}\uparrow,\downarrow{\}}$ are particle density operators, $\mu$ is the particle number chemical potential 
which in this calculation is assumed to be the same for both pairing species. The theory is defined in $d=4-\varepsilon$ spatial dimensions.
After Hubbard-Stratonovich transformation, the Lagrangian density (\ref{lagrangian}) takes the 
form
\ber 
\mathcal{L} = \Psi^\+\left(i\d_t + \frac{\sigma_3\grad^2}{2m}+\mu\sigma_3\right)\Psi
 \nonumber \\ - \frac1{c_0} \phi^*\phi
 + \Psi^\+\sigma_+\Psi\phi + \Psi^\+\sigma_-\Psi\phi^*,
\label{HBlagrangian}
\eer
where $\phi$ is an auxiliary di-fermion field, $\Psi$ is a two-component Nambu-Gor'kov field, $\Psi=(\psi_\uparrow,\psi^\+_\downarrow)^T$,
$\sigma_\pm=\frac12(\sigma_1\pm i\sigma_2),$ with $\sigma_{1,2,3}$ being the Pauli matrices.
In this calculation we will concentrate on the unitary regime, $a=\infty,$ which in the dimensional
regularization corresponds to $1/c_0=0$ \cite{Nishida:2006br}.
Once fluctuations of $\phi$ are identified as propagating bosonic degrees of freedom, the system may be viewed as spin up and spin down 
fermions interacting by exchanging spin 0 mass $2m$ bosons. 
The standard perturbative approach applies \cite{Nishida:2006br,PhysRevA.75.063617}. Extension to the
finite temperature-real time perturbation theory ({\it e.g.}, \cite{Landau10}) is straightforward. 
The fermion-boson coupling for the LO calculations is \cite{Nishida:2006br}
\ber
g = \frac{(8\pi^2\varepsilon)^{1/2}}m.
\label{g}
\eer

Transport coefficients can, in principle, be determined using Kubo relations of linear response theory. However, it has been argued that
using a perturbation theory may require re-summation. The equivalent correct procedure is to solve the appropriately perturbed  
Boltzmann equation in which the collision integral is calculated to a finite order in the perturbation theory \cite{PhysRevD.53.5799}. 
This is the approach used in this work.

Momentum flux tensor of a weakly non-ideal non-relativistic fluid is \cite{Landau10}
{
\ber 
\Pi_{\alpha\beta}={\rm P}\delta_{\alpha\beta}+\rho{\rm V}_{\alpha}{\rm V}_{\beta}-\sigma^{'}_{\alpha\beta},~\alpha,\beta=1..d,
\label{Piab}
\eer}
where ${\rm P}$ is pressure, {${\bf V}({\bf x})$} is the hydrodynamic flow velocity field, $\sigma^{'}_{\alpha\beta}$ is the viscous
stress tensor. For a static shear flow ${\bf \nabla}\cdot{\bf {\rm V}}=0$
\ber
\sigma^{'}_{\alpha\beta}=2{\eta}{\rm V}_{\alpha \beta},
{\rm V}_{\alpha \beta}=\frac12\left(\frac{\partial{\rm V}_{\beta}}{\partial x_{\alpha}}+\frac{\partial{\rm V}_{\alpha}}{\partial x_{\beta}}\right),~{\rm V}_{\alpha \alpha}\equiv0,
\label{sigmaprime}
\eer
where $\eta$ is the (static) shear viscosity.

We have three interacting liquids: spin up and spin down fermion species interact with each other by exchanging 
bosons. In the unpolarized case we are considering here there are two distribution functions $n_{\uparrow}\equiv n_{\downarrow}=n_F,~n_B.$ 
For a static shear flow the transport equations become
\ber
\frac{{\rm d}n_{F}}{{\rm d} t}
=-n_{F}^0(1 - n_{F}^0)
\left(p_{\alpha}p_{\beta}-\frac{\delta_{\alpha \beta}~p^2}{d}\right)\frac{{\rm V}_{\alpha \beta}}{m T}={\rm C}_{F}[n_{F},n_B],
\label{transporteqF}
\eer
\ber
\frac{{\rm d}n_{B}}{{\rm d} t}
=-n_{B}^0(1 + n_{B}^0)
\left(p_{\alpha}p_{\beta}-\frac{\delta_{\alpha \beta}~p^2}{d}\right)\frac{{\rm V}_{\alpha \beta}}{m T}={\rm C}_{B}[n_{F},n_B],
\label{transporteqB}
\eer
where {$n_{F},~n_B$} are the Boson and Fermion distribution functions, respectively, and $C_{F},~C_B$ are the collision integrals, and 
\ber
n^0_{F}&=&\left({\rm exp}({{\rm E}_{F}(\p)}/{T}) + 1\right)^{-1},~
n^0_{B}=\left({\rm exp}({{\rm E}_{B}(\p)}/{T})- 1\right)^{-1},\nonumber \\
{\rm E}_F(\p)&=&\epsp - \mu,~{\rm E}_B(\p)=\epsp/2 - 2\mu,~\epsp=p^2/2 m,
\label{n0FB}
\eer
are the equilibrium Fermion and Boson distribution functions, respectively \cite{Landau10}.
One seeks solutions as
\ber
n_{F}(\p)&=&n_{F}^0-n_{F}^0(1 - n_{F}^0)
\left(p_{\alpha}p_{\beta}-\frac{\delta_{\alpha \beta}}{d}~p^2\right)\frac{{\rm V}_{\alpha \beta}~\chi_{F}(\epsilon_{\p})}{m T^2},\nonumber \\
n_{B}(\p)&=&n_{B}^0-n_{B}^0(1 + n_{B}^0)
\left(p_{\alpha}p_{\beta}-\frac{\delta_{\alpha \beta}}{d}~p^2\right)\frac{{\rm V}_{\alpha \beta}~\chi_{B}(\epsilon_{\p})}{m T^2}.
\label{npert}
\eer
Subsituting (\ref{npert}) into transport equations (\ref{transporteqF}),~(\ref{transporteqB}) and retaining terms linear in $\chi_F,~\chi_B$ one obtains 
two coupled linea integral equations for $\chi_F,~\chi_B$.
Once {$\chi_F,~\chi_B$} have been determined $\eta$ is given by
{
\ber
\eta&=&
\int_{\p}
(n_{F}^0(1-n_{F}^0)2\chi_{F}(\epsilon_{\p})+n_{B}^0(1+n_{B}^0)\chi_{B}(\epsilon_{\p}))\frac{(p_{\alpha}p_{\alpha})^2(1-1/d)}{(d^2 + d - 2) m^2 T^2},\nonumber \\\int_{\p}&\equiv&\int \frac{p^{d-1}}{(2\pi)^d}{\rm d}p\frac{2\pi^{d/2}}{\Gamma(d/2)}.
\label{eta}
\eer}
\section{LO Collision Integrals}
\label{sec:Cs}
It is known that in the finite temperature-real time
formalism the equation of motion for the two point correlation function 
{${\rm G}^{-+}(t_1,{\bf x}_1,t_2,{\bf x}_2)=\pm i \<\psi^{\dagger}_{\alpha}(t_2,{\bf x}_2)\psi_{\alpha}(t_1,{\bf x}_1)\>$}, where  
$\pm$ refers to Fermions and Bosons, respectively, in the quasiclassical, or Thomas-Fermi, regime reduces to the transport equation for the distribution functions
\cite{Landau10}.
The LO collision integrals can then be read off and are given by
{
\ber
{\rm C}_{B}[n_F,n_B]&=&i\Sigma^{-+}_{B}(1+n_B)-i \Sigma_B^{+-}n_B|_{k_0=\epsilon_{\bf k}/2 - 2\mu,}\nonumber \\
{\rm C}_{F}[n_F,n_B]&=&i\Sigma^{-+}_{F}(1-n_F)+i \Sigma_F^{+-}n_F|_{k_0=\epsilon_{\bf k}-\mu,}
\label{CBF}
\eer
} 
where $i\Sigma^{+-}_{B,F}(k_0,{\bf k};n_F,n_B),i\Sigma^{-+}_{B,F}(k_0,{\bf k};n_F,n_B)$ are the boson and fermion self energies \cite{Landau10}.
In Eq. (\ref{CBF}) ${\rm n}$s may depend on time as well as on the phase space coordinates, ${\bf x}$ and ${\bf k},$ which are assumed to be slow varying on the microscopic scales. 

The LO contributions to the collision integral are ${\cal O}(\varepsilon^2).$ This is as expected since the LO contributions to fermion-fermion and fermion-boson scattering 
cross-sections are all
${\cal O}(\varepsilon^2).$ 
We have checked that the ${\cal O}(\varepsilon)$ contributions to both boson and fermion self-energies $i\Sigma_B^{+-}(k_0,\k), i\Sigma_B^{-+}(k_0,\k)$ and 
$i\Sigma_F^{+-}(k_0,\k),i\Sigma_F^{-+}(k_0,\k)$ vanish for 
$k_0=\epsilon_{\k}/2 - 2\mu,~k_0=\epsilon_{\k} - \mu,$ respectively.

\begin{figure}[t]
\centering{
\begin{psfrags}
\psfrag{f}{${\cal{O}}(\varepsilon^2)~\Sigma^{+-}_F$}
\psfrag{b}{${\cal{O}}(\varepsilon^2)~\Sigma^{+-}_B$}
\psfrag{g}{${\rm g}$}
\psfrag{+}{${\rm +}$}
\psfrag{-}{${\rm -}$}
\epsfig{figure=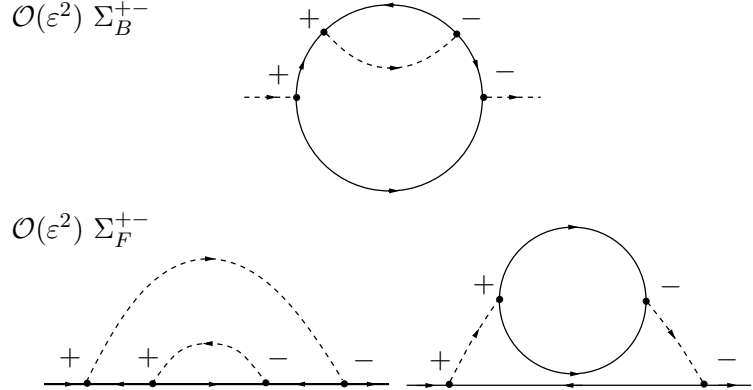, width=.6\textwidth}
\end{psfrags}
}
\caption{Feynman diagrams relevant for the normal phase calculation the collision integrals.
Thin solid lines depict propagating fermions, dashed line is the boson propagator. The boson self-energy diagram with the boson line in the lower part of the fermion loop is not shown.
The $-+$ diagrams are completely analogous. The rest of the ${\cal O}(\varepsilon^2)$ contributions to the self energies produce zero contributions to the collision integrals.
}
\label{fig:graphs2}
\end{figure}

Feynman diagrams relevant for the calculation of the collision integrals in the normal phase are shown in Fig. (\ref{fig:graphs2}).
For the rest of this section we will be distinguishing $\uparrow$ and $\downarrow$ fermion species. This is just for convenience. 
At ${\cal O}(\varepsilon^2)$ the fermion collision integral is
\ber
{\rm C}_{\uparrow,\downarrow}={\rm C}_{\uparrow,\downarrow}^1+{\rm C}_{\uparrow,\downarrow}^2,
\label{CnF}
\eer
where
\ber
{\rm C}_{\uparrow}^1&=&
\frac{g^4 2 (2 m)^d}{(2 \pi)^{2 d-1}}\int {\rm d} \epsilon_{\bf q}~
{\rm d}\Omega_{\p}~{\rm d}\Omega_{\q}\times \nonumber \\ &\times&[n_B({\q+2\k})\left(1+n_B({\p+2{\bar \q}})\right)
n_{\uparrow}({\p+{\bar \q}})(1-n_{\uparrow}({\k}))-\nonumber \\&-&
\left(1+n_B({\q+2\k})\right)n_B({\p+2{\bar \q}})
(1-n_{\uparrow}({\p+{\bar \q}}))n_{\uparrow}({\k})], 
\label{CnF1}
\eer
and
\ber
{\rm C}_{\uparrow}^2&=&
\frac{g^4(2 m)^d}{2(2 \pi)^{2 d-1}}\int {\rm d} \epsilon_{\bf q}~
{\rm d}\Omega_{\p}~{\rm d}\Omega_{\q}\times \nonumber \\ &\times&[n_{\downarrow}({\p+{\bar \q}})n_{\uparrow}({{\bar \q}-\p})
(1-n_{\downarrow}({2\q+\k}))(1-n_{\uparrow}({\k}))-\nonumber \\&-&
(1-n_{\downarrow}({\p+{\bar \q}}))(1-n_{\uparrow}({{\bar \q}-\p}))
n_{\downarrow}({2\q+\k}))n_{\uparrow}({\k})], 
\label{CnF2}
\eer
where $|\p|=|\q|,~{{\bar \q}}=\q+\k.$
Expressions for ${\rm C}_{\downarrow}^1$ and 
${\rm C}_{\downarrow}^2$ are given by formulas (\ref{CnF1}), (\ref{CnF2}) with $\uparrow$ subscripts replaced by 
$\downarrow$ subscripts.
$C^1$ comes from the self energy correction to the fermion propagator, while $C^2$ comes the self energy correction to the boson propagator
in the fermion self energy diagrams shown in Fig. (\ref{fig:graphs2}).
The ${\cal O}(\varepsilon^2)$ contribution to the boson collision integral comes from the self energy corrections to the fermion propagators
in the boson self energy diagram shown in Fig. (\ref{fig:graphs2}). It 
is given by
\ber
{\rm C}_{B}&=&
\frac{g^4 (2 m)^d 2^{d-5}}{(2 \pi)^{2 d-1}}\int {\rm d} \epsilon_{\bf q}
~{\rm d}\Omega_{\p}~{\rm d}\Omega_{\q}
\times \nonumber \\ &\times&
\epsilon_{\bf q}^{d-4}
[(1-n_{\downarrow}({\k/2-\q}))n_{\downarrow}({\p+{\bar \q}})(1+n_{B}({\p+2{\bar \q}}))n_{B}({\k}))-\nonumber \\&-&
n_{\downarrow}({\k/2-\q})(1-n_{\downarrow}({\p+{\bar \q}}))n_{B}({\p+2{\bar \q}})(1+n_{B}({\k}))]+
{\{}\uparrow\leftrightarrow\downarrow{\}}, 
\label{CnB}
\eer
where $|\p|=2|\q|,~{{\bar \q}}=\q+\k/2.$
In the LO expressions above $d=4$ and, so, 
$${\rm d}\Omega_{\p}=\int_0^{2\pi}{\rm d}\phi\int_0^{\pi}{\rm d}\omega~{\rm sin}^2\omega\int_0^{\pi}{\rm d}\theta~{\rm sin}\theta$$ is the four dimensional angular integration. 
Also, we note that collision integrals have the standard gain-loss structure \cite{Landau10}.
Evaluated on the equilibrium distribution functions (\ref{n0FB}) (\ref{CnF1}),(\ref{CnF2}),(\ref{CnB}) vanish, as they should.
\section{Solution of Linearized Transport Equations, Result and Conclusion}
\label{sec:result}
Substituting (\ref{npert}) into (\ref{transporteqF}),~(\ref{transporteqB}) two linear integral equations for $\chi_F(\epsilon_{\bf k}),~\chi_B(\epsilon_{\bf k})$ have been obtained.
The expressions for the linearized integral operators are somewhat bulky and will not be shown here.  

First, we have checked the integrals for several suspected collinear divergences coming from the Bose distribution functions when $\hat{p},~\hat{q}$ and $\hat{k}$ from 
(\ref{CnF1}),~(\ref{CnF2}),~(\ref{CnB}) are (anti)parallel. 
We have observed that in $d=4$ all the suspected divergences produce finite contributions. This has been verified {\it a posteriori} numerically using the solutions obtained. 

The linearized equations have been solved numerically for the hydrodynamic flow with ${\rm V}_{34} \neq 0$ ({\it cf.} \cite{1126-6708-2005-09-076}) by defining the unknown functions on a discretized finite domain. 
Thus the problem has been reduced to the system of linear algebraic equations. However, due to non-locality of the kernels (\ref{CnF1}),~(\ref{CnF2}),~(\ref{CnB}) 
the resulting system of linear equations is not closed, {\it i.e.}, for a fixed domain of the argument $0 \leq \epsilon_{\bf k}/T \leq \Lambda$ the integral operators 
will always depend on values of $\chi_F(\epsilon_{\bf k}),~\chi_B(\epsilon_{\bf k})$ for $\epsilon_{\bf k}/T \geq \Lambda.$
An {\it ad hoc} solution used here is to simply set $\chi_F(\epsilon_{\bf k}),~\chi_B(\epsilon_{\bf k})\equiv 0$ for $\epsilon_{\bf k} \geq \Lambda.$ 
The seven fold integrations have been done numerically by the {\rm Mathematica} software using {\rm Adaptive Monte Carlo} method in the {\rm NIntegrate} function. 
We have checked for convergence of our solutions with respect to both the lattice spacing and the value of the cutoff, $\Lambda.$ 
\begin{figure}[t]
\centering{
\epsfig{figure=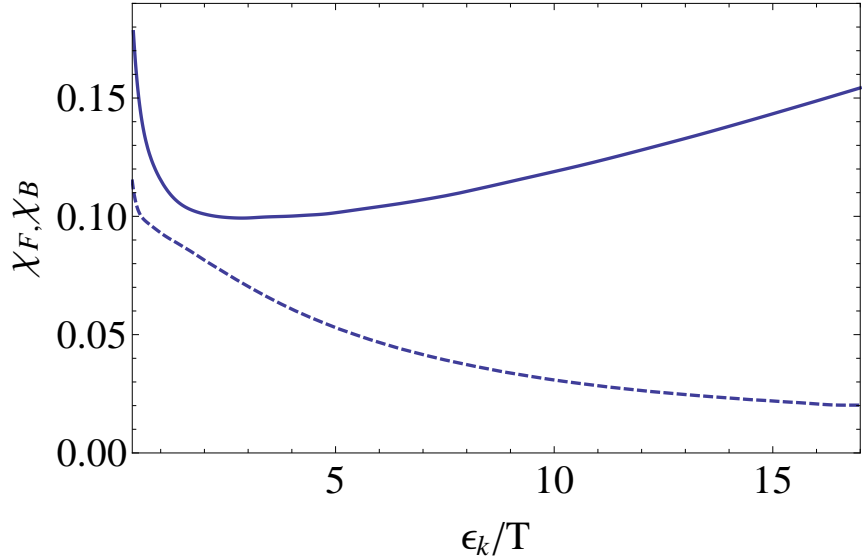, width=.7\textwidth}
}
\caption{Solutions to the linearized transport equations obtained by the method described in text; $\chi_B$ is the solid line,  $\chi_F$ is the dashed line.
}
\label{fig:chiFB}
\end{figure}
The solutions obtained have been converted back into the ``continuous'' form using {\rm Interpolation} function of {\rm Mathematica} and fed back into the linearized transport equations. 
We have found that for $0 \leq \epsilon_{\bf k}/T \leq 13$ relevant to the $\eta$ calculation the ratio of the L.H.S. to the R.H.S. 
was $1.07\pm0.07$ for the linearized Boltzmann equation for fermions
and  $0.97\pm0.03$ for the linearized Boltzmann equation for the boson distribution function.
From this it was concluded that reasonably accurate solutions had been found. The errors may be further reduced by making the lattice finer and the cutoff $\Lambda$ bigger,
but even at this level they do not seem to be the dominant source of uncertainty in the result. Functions $\chi_F(\epsilon_{\bf k}),~\chi_B(\epsilon_{\bf k})$ are plotted in Fig. \ref{fig:chiFB}.
Note that the approach used here differs from the standard variational approach described in the literature where one seeks the solution(s) as an expansion in some basis which results
in the lower bound for the viscosity \cite{Landau10}. For some recent applications of the variational approach see, {\it e.g.}, \cite{1126-6708-2005-09-076,PhysRevA.76.053607}. 

The pressure of the normal phase for $d=4-\varepsilon$ has been computed by Nishida to NLO \cite{PhysRevA.75.063618} 
{
\ber
{\rm P}&=&{\rm T}\left(\frac{m T}{2 \pi}\right)^{d/2}
\left(\frac{11~\zeta(3)}{2}-\frac{9~{\rm ln}2~\zeta(3)+11\zeta^{'}(3)}{4}\varepsilon-C_p\varepsilon+\frac{\pi^2}{6}\frac{\mu}{T}+\frac{2\pi^2}{3}\frac{\mu_B}{T}\right),
\nonumber \\
C_p&\simeq&8.4,~\mu_B=2\mu
\sim\varepsilon.
\label{Pnormal}
\eer}
The entropy density is ${\rm s}=\partial {\rm P}/\partial T.$ 
Evaluating (\ref{eta}) on the linearized transport equations solutions $\chi_F(\epsilon_{\bf k}),~\chi_B(\epsilon_{\bf k})$ and using (\ref{Pnormal}) to LO in $\varepsilon$ we get 
(recall $\hbar=k_B=1$)
\ber
\frac{\eta}{{\rm s}}\simeq\frac{0.11}{\varepsilon^2}(1+{\cal O}(\varepsilon))=\frac{1.44}{\varepsilon^2}\frac{1}{4 \pi}(1+{\cal O}(\varepsilon)).
\label{etas}
\eer
The LO normal phase ratio turns out to be temperature independent.

Here we make the simplest possible extrapolation for $d=3$ by setting $\varepsilon=1$ in (\ref{etas}). We see that
our prediction does not contradict the experimentally obtained bound (\ref{etasexpt}) and is consistent with the conjectured bound 
(\ref{etasbound}). A recent {\it first principles} calculation of the normal phase $\eta/s$ used Kubo linear response approach with the stress 
tensor correlator computed on the lattice \cite{Wlazlowski:2012jb}. The $d=3$ prediction of this work, (\ref{etas}) with $\varepsilon=1,$ appears 
to coincide with the $(\eta/s)_{min}\simeq 0.11$ quoted in \cite{Wlazlowski:2012jb}. We believe this to be a fluke.
Other theoretical approaches yielded $\eta/s$ ratios exceeding the bound (\ref{etasbound}) by, {\it e.g.}, $\sim 7$ \cite{Enss2011770} and 4.7 \cite{LeClair:2010au}.

The main conclusion of this work is that $\varepsilon$-expansion is a versatile useful tool for the unitary Fermi gas already at LO and NLO.
Further investigation aimed at high precision calculations of various observables is fully warranted.
\section{Acknowledgment}
This work has been partially supported by the US Department of Energy grant DE-FG52-08NA28921. The author is grateful to T.~Luu, T.~Ihle, and D.~Kroll for discussions.
\bibliography{cfl2012}

\begin{thebibliography}{46}
\expandafter\ifx\csname natexlab\endcsname\relax\def\natexlab#1{#1}\fi
\expandafter\ifx\csname bibnamefont\endcsname\relax
  \def\bibnamefont#1{#1}\fi
\expandafter\ifx\csname bibfnamefont\endcsname\relax
  \def\bibfnamefont#1{#1}\fi
\expandafter\ifx\csname citenamefont\endcsname\relax
  \def\citenamefont#1{#1}\fi
\expandafter\ifx\csname url\endcsname\relax
  \def\url#1{\texttt{#1}}\fi
\expandafter\ifx\csname urlprefix\endcsname\relax\def\urlprefix{URL }\fi
\providecommand{\bibinfo}[2]{#2}
\providecommand{\eprint}[2][]{\url{#2}}

\bibitem[{\citenamefont{Bertsch}(2001)}]{Bertsch:1999xx}
\bibinfo{author}{\bibfnamefont{G.}~\bibnamefont{Bertsch}},
  \bibinfo{journal}{Int. J. Mod. Phys. B} \textbf{\bibinfo{volume}{15}}
  (\bibinfo{year}{2001}).

\bibitem[{\citenamefont{Yakovlev and Pethick}(2004)}]{Yakovlev:2004iq}
\bibinfo{author}{\bibfnamefont{D.~G.} \bibnamefont{Yakovlev}} \bibnamefont{and}
  \bibinfo{author}{\bibfnamefont{C.~J.} \bibnamefont{Pethick}},
  \bibinfo{journal}{Ann. Rev. Astron. Astrophys.}
  \textbf{\bibinfo{volume}{42}}, \bibinfo{pages}{169} (\bibinfo{year}{2004}),
  \eprint{astro-ph/0402143}.

\bibitem[{\citenamefont{Regal et~al.}(2004)\citenamefont{Regal, Greiner, and
  Jin}}]{PhysRevLett.92.040403}
\bibinfo{author}{\bibfnamefont{C.~A.} \bibnamefont{Regal}},
  \bibinfo{author}{\bibfnamefont{M.}~\bibnamefont{Greiner}}, \bibnamefont{and}
  \bibinfo{author}{\bibfnamefont{D.~S.} \bibnamefont{Jin}},
  \bibinfo{journal}{Phys. Rev. Lett.} \textbf{\bibinfo{volume}{92}},
  \bibinfo{pages}{040403} (\bibinfo{year}{2004}).

\bibitem[{\citenamefont{Bartenstein et~al.}(2004)\citenamefont{Bartenstein,
  Altmeyer, Riedl, Jochim, Chin, Denschlag, and Grimm}}]{PhysRevLett.92.120401}
\bibinfo{author}{\bibfnamefont{M.}~\bibnamefont{Bartenstein}},
  \bibinfo{author}{\bibfnamefont{A.}~\bibnamefont{Altmeyer}},
  \bibinfo{author}{\bibfnamefont{S.}~\bibnamefont{Riedl}},
  \bibinfo{author}{\bibfnamefont{S.}~\bibnamefont{Jochim}},
  \bibinfo{author}{\bibfnamefont{C.}~\bibnamefont{Chin}},
  \bibinfo{author}{\bibfnamefont{J.~H.} \bibnamefont{Denschlag}},
  \bibnamefont{and} \bibinfo{author}{\bibfnamefont{R.}~\bibnamefont{Grimm}},
  \bibinfo{journal}{Phys. Rev. Lett.} \textbf{\bibinfo{volume}{92}},
  \bibinfo{pages}{120401} (\bibinfo{year}{2004}).

\bibitem[{\citenamefont{Zwierlein et~al.}(2004)\citenamefont{Zwierlein, Stan,
  Schunck, Raupach, Kerman, and Ketterle}}]{PhysRevLett.92.120403}
\bibinfo{author}{\bibfnamefont{M.~W.} \bibnamefont{Zwierlein}},
  \bibinfo{author}{\bibfnamefont{C.~A.} \bibnamefont{Stan}},
  \bibinfo{author}{\bibfnamefont{C.~H.} \bibnamefont{Schunck}},
  \bibinfo{author}{\bibfnamefont{S.~M.~F.} \bibnamefont{Raupach}},
  \bibinfo{author}{\bibfnamefont{A.~J.} \bibnamefont{Kerman}},
  \bibnamefont{and} \bibinfo{author}{\bibfnamefont{W.}~\bibnamefont{Ketterle}},
  \bibinfo{journal}{Phys. Rev. Lett.} \textbf{\bibinfo{volume}{92}},
  \bibinfo{pages}{120403} (\bibinfo{year}{2004}).

\bibitem[{\citenamefont{Kinast et~al.}(2004)\citenamefont{Kinast, Hemmer, Gehm,
  Turlapov, and Thomas}}]{PhysRevLett.92.150402}
\bibinfo{author}{\bibfnamefont{J.}~\bibnamefont{Kinast}},
  \bibinfo{author}{\bibfnamefont{S.~L.} \bibnamefont{Hemmer}},
  \bibinfo{author}{\bibfnamefont{M.~E.} \bibnamefont{Gehm}},
  \bibinfo{author}{\bibfnamefont{A.}~\bibnamefont{Turlapov}}, \bibnamefont{and}
  \bibinfo{author}{\bibfnamefont{J.~E.} \bibnamefont{Thomas}},
  \bibinfo{journal}{Phys. Rev. Lett.} \textbf{\bibinfo{volume}{92}},
  \bibinfo{pages}{150402} (\bibinfo{year}{2004}).

\bibitem[{\citenamefont{Bourdel et~al.}(2004)\citenamefont{Bourdel, Khaykovich,
  Cubizolles, Zhang, Chevy, Teichmann, Tarruell, Kokkelmans, and
  Salomon}}]{PhysRevLett.93.050401}
\bibinfo{author}{\bibfnamefont{T.}~\bibnamefont{Bourdel}},
  \bibinfo{author}{\bibfnamefont{L.}~\bibnamefont{Khaykovich}},
  \bibinfo{author}{\bibfnamefont{J.}~\bibnamefont{Cubizolles}},
  \bibinfo{author}{\bibfnamefont{J.}~\bibnamefont{Zhang}},
  \bibinfo{author}{\bibfnamefont{F.}~\bibnamefont{Chevy}},
  \bibinfo{author}{\bibfnamefont{M.}~\bibnamefont{Teichmann}},
  \bibinfo{author}{\bibfnamefont{L.}~\bibnamefont{Tarruell}},
  \bibinfo{author}{\bibfnamefont{S.~J. J. M.~F.} \bibnamefont{Kokkelmans}},
  \bibnamefont{and} \bibinfo{author}{\bibfnamefont{C.}~\bibnamefont{Salomon}},
  \bibinfo{journal}{Phys. Rev. Lett.} \textbf{\bibinfo{volume}{93}},
  \bibinfo{pages}{050401} (\bibinfo{year}{2004}).

\bibitem[{\citenamefont{Bloch et~al.}(2008)\citenamefont{Bloch, Dalibard, and
  Zwerger}}]{bloch-2008-80}
\bibinfo{author}{\bibfnamefont{I.}~\bibnamefont{Bloch}},
  \bibinfo{author}{\bibfnamefont{J.}~\bibnamefont{Dalibard}}, \bibnamefont{and}
  \bibinfo{author}{\bibfnamefont{W.}~\bibnamefont{Zwerger}},
  \bibinfo{journal}{Reviews of Modern Physics} \textbf{\bibinfo{volume}{80}},
  \bibinfo{pages}{885} (\bibinfo{year}{2008}).

\bibitem[{\citenamefont{Bour et~al.}(2011)\citenamefont{Bour, Li, Lee,
  Mei\ss{}ner, and Mitas}}]{PhysRevA.83.063619}
\bibinfo{author}{\bibfnamefont{S.}~\bibnamefont{Bour}},
  \bibinfo{author}{\bibfnamefont{X.}~\bibnamefont{Li}},
  \bibinfo{author}{\bibfnamefont{D.}~\bibnamefont{Lee}},
  \bibinfo{author}{\bibfnamefont{U.-G.} \bibnamefont{Mei\ss{}ner}},
  \bibnamefont{and} \bibinfo{author}{\bibfnamefont{L.}~\bibnamefont{Mitas}},
  \bibinfo{journal}{Phys. Rev. A} \textbf{\bibinfo{volume}{83}},
  \bibinfo{pages}{063619} (\bibinfo{year}{2011}),
  \urlprefix\url{http://link.aps.org/doi/10.1103/PhysRevA.83.063619}.

\bibitem[{\citenamefont{Veillette et~al.}(2007)\citenamefont{Veillette, Sheehy,
  and Radzihovsky}}]{veillette:043614}
\bibinfo{author}{\bibfnamefont{M.~Y.} \bibnamefont{Veillette}},
  \bibinfo{author}{\bibfnamefont{D.~E.} \bibnamefont{Sheehy}},
  \bibnamefont{and}
  \bibinfo{author}{\bibfnamefont{L.}~\bibnamefont{Radzihovsky}},
  \bibinfo{journal}{Physical Review A (Atomic, Molecular, and Optical Physics)}
  \textbf{\bibinfo{volume}{75}}, \bibinfo{eid}{043614}
  (pages~\bibinfo{numpages}{13}) (\bibinfo{year}{2007}),
  \urlprefix\url{http://link.aps.org/abstract/PRA/v75/e043614}.

\bibitem[{\citenamefont{Diehl et~al.}(2007)\citenamefont{Diehl, Gies,
  Pawlowski, and Wetterich}}]{diehl:053627}
\bibinfo{author}{\bibfnamefont{S.}~\bibnamefont{Diehl}},
  \bibinfo{author}{\bibfnamefont{H.}~\bibnamefont{Gies}},
  \bibinfo{author}{\bibfnamefont{J.~M.} \bibnamefont{Pawlowski}},
  \bibnamefont{and}
  \bibinfo{author}{\bibfnamefont{C.}~\bibnamefont{Wetterich}},
  \bibinfo{journal}{Physical Review A (Atomic, Molecular, and Optical Physics)}
  \textbf{\bibinfo{volume}{76}}, \bibinfo{eid}{053627}
  (pages~\bibinfo{numpages}{18}) (\bibinfo{year}{2007}),
  \urlprefix\url{http://link.aps.org/abstract/PRA/v76/e053627}.

\bibitem[{\citenamefont{Bulgac}(2007)}]{bulgac:040502}
\bibinfo{author}{\bibfnamefont{A.}~\bibnamefont{Bulgac}},
  \bibinfo{journal}{Physical Review A (Atomic, Molecular, and Optical Physics)}
  \textbf{\bibinfo{volume}{76}}, \bibinfo{eid}{040502}
  (pages~\bibinfo{numpages}{4}) (\bibinfo{year}{2007}),
  \urlprefix\url{http://link.aps.org/abstract/PRA/v76/e040502}.

\bibitem[{\citenamefont{Giorgini et~al.}(2008)\citenamefont{Giorgini,
  Pitaevskii, and Stringari}}]{giorgini:1215}
\bibinfo{author}{\bibfnamefont{S.}~\bibnamefont{Giorgini}},
  \bibinfo{author}{\bibfnamefont{L.~P.} \bibnamefont{Pitaevskii}},
  \bibnamefont{and}
  \bibinfo{author}{\bibfnamefont{S.}~\bibnamefont{Stringari}},
  \bibinfo{journal}{Reviews of Modern Physics} \textbf{\bibinfo{volume}{80}},
  \bibinfo{eid}{1215} (\bibinfo{year}{2008}),
  \urlprefix\url{http://link.aps.org/abstract/RMP/v80/p1215}.

\bibitem[{\citenamefont{Nussinov and Nussinov}(2004)}]{Nussinov:2004nn}
\bibinfo{author}{\bibfnamefont{Z.}~\bibnamefont{Nussinov}} \bibnamefont{and}
  \bibinfo{author}{\bibfnamefont{S.}~\bibnamefont{Nussinov}}
  (\bibinfo{year}{2004}), \eprint{cond-mat/0410597}.

\bibitem[{\citenamefont{Wilson and Kogut}(1974)}]{Wilson197475}
\bibinfo{author}{\bibfnamefont{K.~G.} \bibnamefont{Wilson}} \bibnamefont{and}
  \bibinfo{author}{\bibfnamefont{J.}~\bibnamefont{Kogut}},
  \bibinfo{journal}{Physics Reports} \textbf{\bibinfo{volume}{12}},
  \bibinfo{pages}{75 } (\bibinfo{year}{1974}).

\bibitem[{\citenamefont{Nishida and Son}(2006)}]{Nishida:2006br}
\bibinfo{author}{\bibfnamefont{Y.}~\bibnamefont{Nishida}} \bibnamefont{and}
  \bibinfo{author}{\bibfnamefont{D.~T.} \bibnamefont{Son}},
  \bibinfo{journal}{Phys. Rev. Lett.} \textbf{\bibinfo{volume}{97}},
  \bibinfo{pages}{050403} (\bibinfo{year}{2006}), \eprint{cond-mat/0604500}.

\bibitem[{\citenamefont{Chen and Nakano}(2007)}]{Chen:2006wx}
\bibinfo{author}{\bibfnamefont{J.-W.} \bibnamefont{Chen}} \bibnamefont{and}
  \bibinfo{author}{\bibfnamefont{E.}~\bibnamefont{Nakano}},
  \bibinfo{journal}{Phys. Rev.} \textbf{\bibinfo{volume}{A75}},
  \bibinfo{pages}{043620} (\bibinfo{year}{2007}), \eprint{cond-mat/0610011}.

\bibitem[{\citenamefont{Nishida and Son}(2007)}]{PhysRevA.75.063617}
\bibinfo{author}{\bibfnamefont{Y.}~\bibnamefont{Nishida}} \bibnamefont{and}
  \bibinfo{author}{\bibfnamefont{D.~T.} \bibnamefont{Son}},
  \bibinfo{journal}{Phys. Rev. A} \textbf{\bibinfo{volume}{75}},
  \bibinfo{pages}{063617} (\bibinfo{year}{2007}),
  \urlprefix\url{http://link.aps.org/doi/10.1103/PhysRevA.75.063617}.

\bibitem[{\citenamefont{Zinn-Justin}(2002)}]{ZinnJ}
\bibinfo{author}{\bibfnamefont{J.}~\bibnamefont{Zinn-Justin}},
  \emph{\bibinfo{title}{Quantum Field Theory and Critical Phenomena}}
  (\bibinfo{publisher}{Clarendon Press}, \bibinfo{address}{Oxford},
  \bibinfo{year}{2002}), \bibinfo{edition}{2nd} ed.

\bibitem[{\citenamefont{Arnold et~al.}(2007)\citenamefont{Arnold, Drut, and
  Son}}]{arnold-2007-75}
\bibinfo{author}{\bibfnamefont{P.}~\bibnamefont{Arnold}},
  \bibinfo{author}{\bibfnamefont{J.~E.} \bibnamefont{Drut}}, \bibnamefont{and}
  \bibinfo{author}{\bibfnamefont{D.~T.} \bibnamefont{Son}},
  \bibinfo{journal}{Physical Review A} \textbf{\bibinfo{volume}{75}},
  \bibinfo{pages}{043605} (\bibinfo{year}{2007}),
  \urlprefix\url{doi:10.1103/PhysRevA.75.043605}.

\bibitem[{\citenamefont{Nishida}(2009)}]{PhysRevA.79.013627}
\bibinfo{author}{\bibfnamefont{Y.}~\bibnamefont{Nishida}},
  \bibinfo{journal}{Phys. Rev. A} \textbf{\bibinfo{volume}{79}},
  \bibinfo{pages}{013627} (\bibinfo{year}{2009}),
  \urlprefix\url{http://link.aps.org/doi/10.1103/PhysRevA.79.013627}.

\bibitem[{\citenamefont{Forbes et~al.}(2011)\citenamefont{Forbes, Gandolfi, and
  Gezerlis}}]{PhysRevLett.106.235303}
\bibinfo{author}{\bibfnamefont{M.~M.} \bibnamefont{Forbes}},
  \bibinfo{author}{\bibfnamefont{S.}~\bibnamefont{Gandolfi}}, \bibnamefont{and}
  \bibinfo{author}{\bibfnamefont{A.}~\bibnamefont{Gezerlis}},
  \bibinfo{journal}{Phys. Rev. Lett.} \textbf{\bibinfo{volume}{106}},
  \bibinfo{pages}{235303} (\bibinfo{year}{2011}),
  \urlprefix\url{http://link.aps.org/doi/10.1103/PhysRevLett.106.235303}.

\bibitem[{\citenamefont{Bulgac et~al.}(2008)\citenamefont{Bulgac, Drut, and
  Magierski}}]{PhysRevA.78.023625}
\bibinfo{author}{\bibfnamefont{A.}~\bibnamefont{Bulgac}},
  \bibinfo{author}{\bibfnamefont{J.~E.} \bibnamefont{Drut}}, \bibnamefont{and}
  \bibinfo{author}{\bibfnamefont{P.}~\bibnamefont{Magierski}},
  \bibinfo{journal}{Phys. Rev. A} \textbf{\bibinfo{volume}{78}},
  \bibinfo{pages}{023625} (\bibinfo{year}{2008}),
  \urlprefix\url{http://link.aps.org/doi/10.1103/PhysRevA.78.023625}.

\bibitem[{\citenamefont{Endres et~al.}(2012)\citenamefont{Endres, Kaplan, Lee,
  and Nicholson}}]{Endres:2012cw}
\bibinfo{author}{\bibfnamefont{M.~G.} \bibnamefont{Endres}},
  \bibinfo{author}{\bibfnamefont{D.~B.} \bibnamefont{Kaplan}},
  \bibinfo{author}{\bibfnamefont{J.-W.} \bibnamefont{Lee}}, \bibnamefont{and}
  \bibinfo{author}{\bibfnamefont{A.~N.} \bibnamefont{Nicholson}}
  (\bibinfo{year}{2012}), \eprint{1203.3169}.

\bibitem[{\citenamefont{Carlson and Reddy}(2005)}]{PhysRevLett.95.060401}
\bibinfo{author}{\bibfnamefont{J.}~\bibnamefont{Carlson}} \bibnamefont{and}
  \bibinfo{author}{\bibfnamefont{S.}~\bibnamefont{Reddy}},
  \bibinfo{journal}{Phys. Rev. Lett.} \textbf{\bibinfo{volume}{95}},
  \bibinfo{pages}{060401} (\bibinfo{year}{2005}).

\bibitem[{\citenamefont{Nishida and Son}(2010)}]{Nishida:2010tm}
\bibinfo{author}{\bibfnamefont{Y.}~\bibnamefont{Nishida}} \bibnamefont{and}
  \bibinfo{author}{\bibfnamefont{D.~T.} \bibnamefont{Son}}
  (\bibinfo{year}{2010}), \eprint{1004.3597}.

\bibitem[{\citenamefont{Rupak}(2006)}]{Rupak:2006jj}
\bibinfo{author}{\bibfnamefont{G.}~\bibnamefont{Rupak}} (\bibinfo{year}{2006}),
  \eprint{nucl-th/0605074}.

\bibitem[{\citenamefont{Kryjevski}(2010)}]{PhysRevA.81.043617}
\bibinfo{author}{\bibfnamefont{A.}~\bibnamefont{Kryjevski}},
  \bibinfo{journal}{Phys. Rev. A} \textbf{\bibinfo{volume}{81}},
  \bibinfo{pages}{043617} (\bibinfo{year}{2010}).

\bibitem[{\citenamefont{Rupak et~al.}(2007)\citenamefont{Rupak, Schafer, and
  Kryjevski}}]{Rupak:2006et}
\bibinfo{author}{\bibfnamefont{G.}~\bibnamefont{Rupak}},
  \bibinfo{author}{\bibfnamefont{T.}~\bibnamefont{Schafer}}, \bibnamefont{and}
  \bibinfo{author}{\bibfnamefont{A.}~\bibnamefont{Kryjevski}},
  \bibinfo{journal}{Phys. Rev.} \textbf{\bibinfo{volume}{A75}},
  \bibinfo{pages}{023606} (\bibinfo{year}{2007}), \eprint{cond-mat/0607834}.

\bibitem[{\citenamefont{Kryjevski}(2008)}]{Kryjevski:2007au}
\bibinfo{author}{\bibfnamefont{A.}~\bibnamefont{Kryjevski}},
  \bibinfo{journal}{Phys. Rev.} \textbf{\bibinfo{volume}{A78}},
  \bibinfo{pages}{043610} (\bibinfo{year}{2008}), \eprint{0712.2093}.

\bibitem[{\citenamefont{Nishida}(2007{\natexlab{a}})}]{Nishida:2006rp}
\bibinfo{author}{\bibfnamefont{Y.}~\bibnamefont{Nishida}},
  \bibinfo{journal}{Phys. Rev.} \textbf{\bibinfo{volume}{A75}},
  \bibinfo{pages}{063618} (\bibinfo{year}{2007}{\natexlab{a}}),
  \eprint{cond-mat/0608321}.

\bibitem[{\citenamefont{Cao et~al.}(2011)\citenamefont{Cao, Elliott, Joseph,
  Wu, Petricka, Schäfer, and Thomas}}]{Cao07012011}
\bibinfo{author}{\bibfnamefont{C.}~\bibnamefont{Cao}},
  \bibinfo{author}{\bibfnamefont{E.}~\bibnamefont{Elliott}},
  \bibinfo{author}{\bibfnamefont{J.}~\bibnamefont{Joseph}},
  \bibinfo{author}{\bibfnamefont{H.}~\bibnamefont{Wu}},
  \bibinfo{author}{\bibfnamefont{J.}~\bibnamefont{Petricka}},
  \bibinfo{author}{\bibfnamefont{T.}~\bibnamefont{Schäfer}}, \bibnamefont{and}
  \bibinfo{author}{\bibfnamefont{J.~E.} \bibnamefont{Thomas}},
  \bibinfo{journal}{Science} \textbf{\bibinfo{volume}{331}},
  \bibinfo{pages}{58} (\bibinfo{year}{2011}),
  \eprint{http://www.sciencemag.org/content/331/6013/58.full.pdf},
  \urlprefix\url{http://www.sciencemag.org/content/331/6013/58.abstract}.

\bibitem[{\citenamefont{Schäfer and Teaney}(2009)}]{0034-4885-72-12-126001}
\bibinfo{author}{\bibfnamefont{T.}~\bibnamefont{Schäfer}} \bibnamefont{and}
  \bibinfo{author}{\bibfnamefont{D.}~\bibnamefont{Teaney}},
  \bibinfo{journal}{Reports on Progress in Physics}
  \textbf{\bibinfo{volume}{72}}, \bibinfo{pages}{126001}
  (\bibinfo{year}{2009}),
  \urlprefix\url{http://stacks.iop.org/0034-4885/72/i=12/a=126001}.

\bibitem[{\citenamefont{Kovtun et~al.}(2005)\citenamefont{Kovtun, Son, and
  Starinets}}]{Kovtun:2004de}
\bibinfo{author}{\bibfnamefont{P.}~\bibnamefont{Kovtun}},
  \bibinfo{author}{\bibfnamefont{D.~T.} \bibnamefont{Son}}, \bibnamefont{and}
  \bibinfo{author}{\bibfnamefont{A.~O.} \bibnamefont{Starinets}},
  \bibinfo{journal}{Phys. Rev. Lett.} \textbf{\bibinfo{volume}{94}},
  \bibinfo{pages}{111601} (\bibinfo{year}{2005}), \eprint{hep-th/0405231}.

\bibitem[{\citenamefont{Dusling and Teaney}(2008)}]{PhysRevC.77.034905}
\bibinfo{author}{\bibfnamefont{K.}~\bibnamefont{Dusling}} \bibnamefont{and}
  \bibinfo{author}{\bibfnamefont{D.}~\bibnamefont{Teaney}},
  \bibinfo{journal}{Phys. Rev. C} \textbf{\bibinfo{volume}{77}},
  \bibinfo{pages}{034905} (\bibinfo{year}{2008}),
  \urlprefix\url{http://link.aps.org/doi/10.1103/PhysRevC.77.034905}.

\bibitem[{\citenamefont{Romatschke and
  Romatschke}(2007)}]{PhysRevLett.99.172301}
\bibinfo{author}{\bibfnamefont{P.}~\bibnamefont{Romatschke}} \bibnamefont{and}
  \bibinfo{author}{\bibfnamefont{U.}~\bibnamefont{Romatschke}},
  \bibinfo{journal}{Phys. Rev. Lett.} \textbf{\bibinfo{volume}{99}},
  \bibinfo{pages}{172301} (\bibinfo{year}{2007}),
  \urlprefix\url{http://link.aps.org/doi/10.1103/PhysRevLett.99.172301}.

\bibitem[{\citenamefont{Sch\"afer}(2007)}]{PhysRevA.76.063618}
\bibinfo{author}{\bibfnamefont{T.}~\bibnamefont{Sch\"afer}},
  \bibinfo{journal}{Phys. Rev. A} \textbf{\bibinfo{volume}{76}},
  \bibinfo{pages}{063618} (\bibinfo{year}{2007}),
  \urlprefix\url{http://link.aps.org/doi/10.1103/PhysRevA.76.063618}.

\bibitem[{\citenamefont{Turlapov et~al.}(2008)\citenamefont{Turlapov, Kinast,
  Clancy, Luo, Joseph, and Thomas}}]{springerlink:10.1007/s10909-007-9589-1}
\bibinfo{author}{\bibfnamefont{A.}~\bibnamefont{Turlapov}},
  \bibinfo{author}{\bibfnamefont{J.}~\bibnamefont{Kinast}},
  \bibinfo{author}{\bibfnamefont{B.}~\bibnamefont{Clancy}},
  \bibinfo{author}{\bibfnamefont{L.}~\bibnamefont{Luo}},
  \bibinfo{author}{\bibfnamefont{J.}~\bibnamefont{Joseph}}, \bibnamefont{and}
  \bibinfo{author}{\bibfnamefont{J.}~\bibnamefont{Thomas}},
  \bibinfo{journal}{Journal of Low Temperature Physics}
  \textbf{\bibinfo{volume}{150}}, \bibinfo{pages}{567} (\bibinfo{year}{2008}),
  ISSN \bibinfo{issn}{0022-2291}, \bibinfo{note}{10.1007/s10909-007-9589-1},
  \urlprefix\url{http://dx.doi.org/10.1007/s10909-007-9589-1}.

\bibitem[{\citenamefont{Lifshitz and Pitaevskii}(1981)}]{Landau10}
\bibinfo{author}{\bibfnamefont{E.~M.} \bibnamefont{Lifshitz}} \bibnamefont{and}
  \bibinfo{author}{\bibfnamefont{L.~P.} \bibnamefont{Pitaevskii}},
  \emph{\bibinfo{title}{Physical Kinetics}} (\bibinfo{publisher}{Pergamon
  Press}, \bibinfo{address}{New York}, \bibinfo{year}{1981}),
  \bibinfo{edition}{1st} ed.

\bibitem[{\citenamefont{Jeon and Yaffe}(1996)}]{PhysRevD.53.5799}
\bibinfo{author}{\bibfnamefont{S.}~\bibnamefont{Jeon}} \bibnamefont{and}
  \bibinfo{author}{\bibfnamefont{L.~G.} \bibnamefont{Yaffe}},
  \bibinfo{journal}{Phys. Rev. D} \textbf{\bibinfo{volume}{53}},
  \bibinfo{pages}{5799} (\bibinfo{year}{1996}),
  \urlprefix\url{http://link.aps.org/doi/10.1103/PhysRevD.53.5799}.

\bibitem[{\citenamefont{Manuel et~al.}(2005)\citenamefont{Manuel, Dobado, and
  Llanes-Estrada}}]{1126-6708-2005-09-076}
\bibinfo{author}{\bibfnamefont{C.}~\bibnamefont{Manuel}},
  \bibinfo{author}{\bibfnamefont{A.}~\bibnamefont{Dobado}}, \bibnamefont{and}
  \bibinfo{author}{\bibfnamefont{F.~J.} \bibnamefont{Llanes-Estrada}},
  \bibinfo{journal}{Journal of High Energy Physics}
  \textbf{\bibinfo{volume}{2005}}, \bibinfo{pages}{076} (\bibinfo{year}{2005}),
  \urlprefix\url{http://stacks.iop.org/1126-6708/2005/i=09/a=076}.

\bibitem[{\citenamefont{Rupak and Sch\"afer}(2007)}]{PhysRevA.76.053607}
\bibinfo{author}{\bibfnamefont{G.}~\bibnamefont{Rupak}} \bibnamefont{and}
  \bibinfo{author}{\bibfnamefont{T.}~\bibnamefont{Sch\"afer}},
  \bibinfo{journal}{Phys. Rev. A} \textbf{\bibinfo{volume}{76}},
  \bibinfo{pages}{053607} (\bibinfo{year}{2007}),
  \urlprefix\url{http://link.aps.org/doi/10.1103/PhysRevA.76.053607}.

\bibitem[{\citenamefont{Nishida}(2007{\natexlab{b}})}]{PhysRevA.75.063618}
\bibinfo{author}{\bibfnamefont{Y.}~\bibnamefont{Nishida}},
  \bibinfo{journal}{Phys. Rev. A} \textbf{\bibinfo{volume}{75}},
  \bibinfo{pages}{063618} (\bibinfo{year}{2007}{\natexlab{b}}),
  \urlprefix\url{http://link.aps.org/doi/10.1103/PhysRevA.75.063618}.

\bibitem[{\citenamefont{Wlazlowski et~al.}(2012)\citenamefont{Wlazlowski,
  Magierski, and Drut}}]{Wlazlowski:2012jb}
\bibinfo{author}{\bibfnamefont{G.}~\bibnamefont{Wlazlowski}},
  \bibinfo{author}{\bibfnamefont{P.}~\bibnamefont{Magierski}},
  \bibnamefont{and} \bibinfo{author}{\bibfnamefont{J.~E.} \bibnamefont{Drut}}
  (\bibinfo{year}{2012}), \eprint{1204.0270}.

\bibitem[{\citenamefont{Enss et~al.}(2011)\citenamefont{Enss, Haussmann, and
  Zwerger}}]{Enss2011770}
\bibinfo{author}{\bibfnamefont{T.}~\bibnamefont{Enss}},
  \bibinfo{author}{\bibfnamefont{R.}~\bibnamefont{Haussmann}},
  \bibnamefont{and} \bibinfo{author}{\bibfnamefont{W.}~\bibnamefont{Zwerger}},
  \bibinfo{journal}{Annals of Physics} \textbf{\bibinfo{volume}{326}},
  \bibinfo{pages}{770 } (\bibinfo{year}{2011}), ISSN \bibinfo{issn}{0003-4916},
  \urlprefix\url{http://www.sciencedirect.com/science/article/pii/S00034916100%
0179X}.

\bibitem[{\citenamefont{LeClair}(2011)}]{LeClair:2010au}
\bibinfo{author}{\bibfnamefont{A.}~\bibnamefont{LeClair}},
  \bibinfo{journal}{New J.Phys.} \textbf{\bibinfo{volume}{13}},
  \bibinfo{pages}{055015} (\bibinfo{year}{2011}), \eprint{1012.5653}.

\end{thebibliography}
\end{document}